\documentclass[a4paper,fleqn,usenatbib]{mnras}

\usepackage[T1]{fontenc}
\usepackage{ae,aecompl}

\usepackage{graphicx}	
\usepackage{amsmath}	
\usepackage{amssymb}	
\usepackage{multirow}   
\usepackage{txfonts}

\title[Two-component debris discs and q$^1$~Eri]{Origin and evolution of two-component debris discs \\ and an application to the q$^1$~Eridani system}

\author[Ch. Sch\"uppler et al.]{Christian~Sch\"uppler,$^{1}$\thanks{E-mail: christian.schueppler@uni-jena.de}
Alexander~V.~Krivov,$^{1}$
Torsten~L\"ohne,$^{1}$
Mark Booth,$^{1,2}$
\newauthor
Florian Kirchschlager,$^{3}$
and Sebastian~Wolf$^{3}$
\\
$^{1}$Astrophysikalisches Institut und Universit\"atssternwarte, 
           Friedrich-Schiller-Universit\"at Jena, \\
           Schillerg\"a{\ss}chen~2--3, 07745 Jena, 
           Germany\\
$^{2}$Instituto de Astrof\'isica, Pontificia Universidad Cat\'olica de Chile, Vicu\~na Mackenna 4860, 7820436 Macul, Santiago, Chile\\
$^{3}$Institut f\"ur Theoretische Physik und Astrophysik, Christian-Albrechts-Universit\"at zu Kiel, Leibnizstra\ss{}e 15, 24118 Kiel, Germany
}

\date{Accepted XXX. Received YYY; in original form ZZZ}

\pubyear{2016}

\begin{document}
\label{firstpage}
\pagerange{\pageref{firstpage}--\pageref{lastpage}}
\maketitle

\begin{abstract}
Many debris discs reveal a two-component structure, with an outer Kuiper-belt analogue
and a warm inner component whose origin is still a matter of debate. One possibility is
that warm emission stems from an ``asteroid belt'' closer in to the star. 
We consider a scenario in which a set of giant planets is formed in
an initially extended planetesimal disc. These planets carve a broad gap around their orbits,
splitting up the disc into the outer and the inner belts. After the gas dispersal, both belts
undergo collisional evolution in a steady-state regime.
This scenario is explored with detailed collisional simulations involving realistic physics to
describe a long-term collisional depletion of the two-component disc.
We find that the inner disc may be able to retain larger amounts of material at older ages
than thought before on the basis of simplified analytic models.
We show that the proposed scenario is consistent with a suite of thermal emission and scattered light 
observational data for a bright two-temperature debris disc around a nearby solar-type star 
q$^1$~Eridani. This implies a Solar System-like architecture of the system, with an outer
massive ``Kuiper belt'', an inner ``asteroid belt'', and a few Neptune- to Jupiter-mass
planets in between.
\end{abstract}




\section{Introduction}
\label{sec:intro}
Debris discs, dusty belts of planetesimals around stars, are natural
by-products of planet formation.  
While far-infrared observations of the past decades typically revealed cold, 
distant dust stemming from Kuiper belt analogues, 
recent observations and analyses demonstrate that many discs
possess a rich radial structure. 
There are systems such as Fomalhaut and Vega where dust appears to be present 
all the way through from more than a hundred au to the sublimation 
zone at a few stellar radii
\citep[e.g.,][]{Su2013,Lebreton2013}.

The presence of material in the cavities of the main, 
cold discs seems to be a rule rather than an exception.
Indeed, about two-thirds of debris disc systems
around late-type stars may exhibit an additional warm component
\citep[e.g.,][]{Morales2011,Ballering2013,Chen2014,Pawellek2014,Kennedy&Wyatt2014}.
The nature of that warm component remains a matter of debate.
One possibility is to attribute the warm dust emission to an inner planetesimal belt, 
i.e. to an asteroid belt analogue.
By analogy with the Solar System with its Kuiper belt and asteroid belt, 
such a two-component structure could be created by a set of giant planets.
If these succeeded to form in the disc, they would carve a hole in the protoplanetary disc by 
removing neighbouring dust and gas \citep[][and references therein]{Dipierro2016}.
After the gas dispersal, nascent planets would swiftly remove planetesimals from their chaotic zones.
At the same time, these planets would dynamically excite planetesimals in the zones bracketing 
the planetary region, preventing their further growth to full-size planets.
All this would generate a broad gap in the planetary region, 
splitting up the disc into two distinct debris belts.
So far, the presumed inner belt may have been marginally resolved for two systems, 
$\varepsilon$~Eri \citep{Greaves2014} and HD~107146 \citep{Ricci2015a}.

Instead of being locally produced in an ``asteroid belt'',
warm dust can be transported inward from outer production zones 
by Poynting-Robertson and stellar wind drag.
For low-density systems where collisional timescales are longer than the one 
for transport, the inner disc region is filled by dust, leading to a nearly 
uniform density profile \citep{Wyatt2005,Kennedy&Piette2015}. 
This scenario is able to explain the observed emission of
discs around late-type stars such as $\varepsilon$~Eri (K2~V), HIP~17439 (K2~V), and AU~Mic (M1~V) 
as shown previously by means of collisional modelling  
\citep{Reidemeister2011, Schueppler2014, Schueppler2015}.
However, this may not be a viable alternative for discs around earlier-type stars
where no strong stellar winds are expected.

In this paper, we consider the debris disc around the nearby F-star q$^\mathrm{1}$~Eridani.
Among the discs observed with \textit{Spitzer} and \textit{Herschel},
the q$^\mathrm{1}$~Eri system is outstanding for having a strong infrared excess with 
a fractional luminosity (infrared luminosity of the disc divided by the stellar luminosity) 
of about $(2-3)\times10^{-4}$ \citep{Trilling2008, Eiroa2013}.
The disc has been spatially resolved in the far infrared
\citep{Liseau2008, Liseau2010} and in the scattered light (K.~Stapelfeldt, priv.~comm.).
The images reveal a bright disc with a radius of \mbox{$\sim100$\,au}.
The mid-infrared \textit{Spitzer}/IRS spectrum between 20 and 30\,\micron{} \citep{Chen2006}
clearly hints at the presence of an additional warm debris component, which 
is unresolved in the images.
The fractional luminosity of the warm component is \mbox{$\sim10^{-4}$}, which is only 
a factor of several lower than that of the main disc.
Attempts to reproduce the data with a two-component model 
require the warm dust to be located at several au from the star
\citep{Tanner2009,Pawellek2014}.
Apart from the two-component disc, a Jupiter-mass planet with a semi-major axis of 2\,au 
and an eccentricity of 0.15 has been detected by radial velocity measurements \citep{Butler2006,Marmier2013}.
The region inside the planetary orbit seems to be largely free of dust
as there is no detection of near-infrared excess, 
associated with the presence of hot exozodiacal dust \citep{Ertel2014_exozodis}. 

Explaining the architecture of the q$^\mathrm{1}$~Eri disc is a challenge.
Since q$^\mathrm{1}$~Eri is an F~star, we do not expect strong stellar winds, and the 
Poynting-Robertson drag alone would be inefficient in delivering an amount of material 
sufficient to explain the warm dust emission. 
This might reinforce the planetary scenario described above, 
in which the system would be reminiscent of the Solar System's two-belt architecture
with a set of invisible planets between the belts. 
However, this explanation is questionable, too.
It is well known that the fractional luminosity of a debris belt collisionally evolving in a steady-state 
regime cannot exceed a certain value for a given system's age \citep{Wyatt2007b,Loehne2008}.
That limit is more stringent for older systems and 
for belts closer to the star where the collisional decay is faster. 
Assuming plausible parameters, the model by \citet{Wyatt2007b} suggests that the warm dust in 
the q$^\mathrm{1}$~Eri disc may be indeed too bright, and the system too old, to be compatible with a 
steady-state collisional cascade in an ``asteroid belt''. 
If true, this would probably leave us with a possibility that the observed warm dust 
is an aftermath of a recent major collision between two big planetesimals. 
However, this comes with the caveat that such events in gigayear-old systems may not be 
likely either.

In this paper, we show that the Solar System-like planetary scenario may work, contrary to 
what simplified models of the long-term collisional decay may suggest.
We present a model to explain the formation of both components of the q$^\mathrm{1}$~Eri disc 
self-consistently.
To this end, we simulate the collisional evolution 
of an initially extended planetesimal distribution, 
assumed to be typical for the end stage of a protoplanetary disc,
and let it evolve over the age of the q$^\mathrm{1}$~Eri system.
Our simulations predict the radial and size distribution of the produced 
dust as a function of time, from which thermal emission properties are calculated to 
compare with observations.
The goal is to see whether a sufficient amount of material can survive in the inner region
of the system against collisional depletion to reproduce the available observational data.

Section~\ref{sec:model} presents the long-term collisional evolution of fiducial debris discs.
In Sect.~\ref{sec:application}, we give an overview of the thermal emission data, 
obtained for the q$^\mathrm{1}$~Eri system, and
apply the collisional modelling to the q$^\mathrm{1}$~Eri disc.
Section~\ref{sec:discussion} contains a discussion and Sect.~\ref{sec:conclusions} lists our conclusions.

\section{Long-term collisional evolution model}
\label{sec:model}

\subsection{Modelling method}
\label{sec:method}
We perform the modelling with the \texttt{ACE} code \citep{Krivov2006, Loehne2012, Krivov2013}, 
which simulates the collisional evolution of debris discs.
Starting from an assumed initial distribution of planetesimals, 
the code calculates the production and loss of material in a collisional
cascade by solving the Smoluchowski-Boltzmann kinetic equation.
Radiative and corpuscular forces acting on 
the disc particles are also included. 
\texttt{ACE} provides coupled spatial and size distributions of the objects within the disc, 
ranging from $\sim100$\,km-sized planetesimals down to submicron-sized dust grains.

The code uses object masses, pericentric distances, and eccentricities 
as phase-space variables and assumes the disc to be rotationally symmetric.
Particle densities are averaged over the inclination, from zero to the 
disc's semi-opening angle, which is fixed to half the 
maximum eccentricity of the planetesimals.
The outcome of a two-particle collision is 
controlled by the specific energy 
threshold for disruption and dispersal.
We use the specific disruption energy, $Q_\mathrm{D}^\ast$, 
as expressed in equation~(7) of \cite{Loehne2012}, 
which is a sum of two power laws for the particle size ranges
where $Q_\mathrm{D}^\ast$ is dominated by tensile strength
and self-gravity.
Following the values of basalt presented in \cite{Benz1999},
we assume 
$Q_\mathrm{D,s}=Q_\mathrm{D,g}=5\times10^6~\mathrm{erg}\,\mathrm{g}^{-1}$, 
$b_\mathrm{s}=-0.37$, and $b_\mathrm{g}=1.38$.
If the specific impact energy of a collision exceeds the threshold $Q_\mathrm{D}^\ast$, 
the target breaks apart and the collision is treated as disruptive. 
Beside disruptive collisions, the code also considers cratering, where 
the target is cratered only while the projectile is disrupted, and bouncing collisions, 
where the target stays intact and the projectile is cratered.
The mass distribution of fragments for each collision is assumed 
to follow $m^{-1.83}$, which is the equilibrium value averaged 
over many collisions \citep{Thebault2007}.
Dust transport within the q$^\mathrm{1}$~Eri disc is assumed to play a minor role.
Hence, we do not 
take Poynting-Robertson and stellar wind drag into account 
in the simulations presented in this paper.

\begin{figure*}
\includegraphics[width=0.95\textwidth]{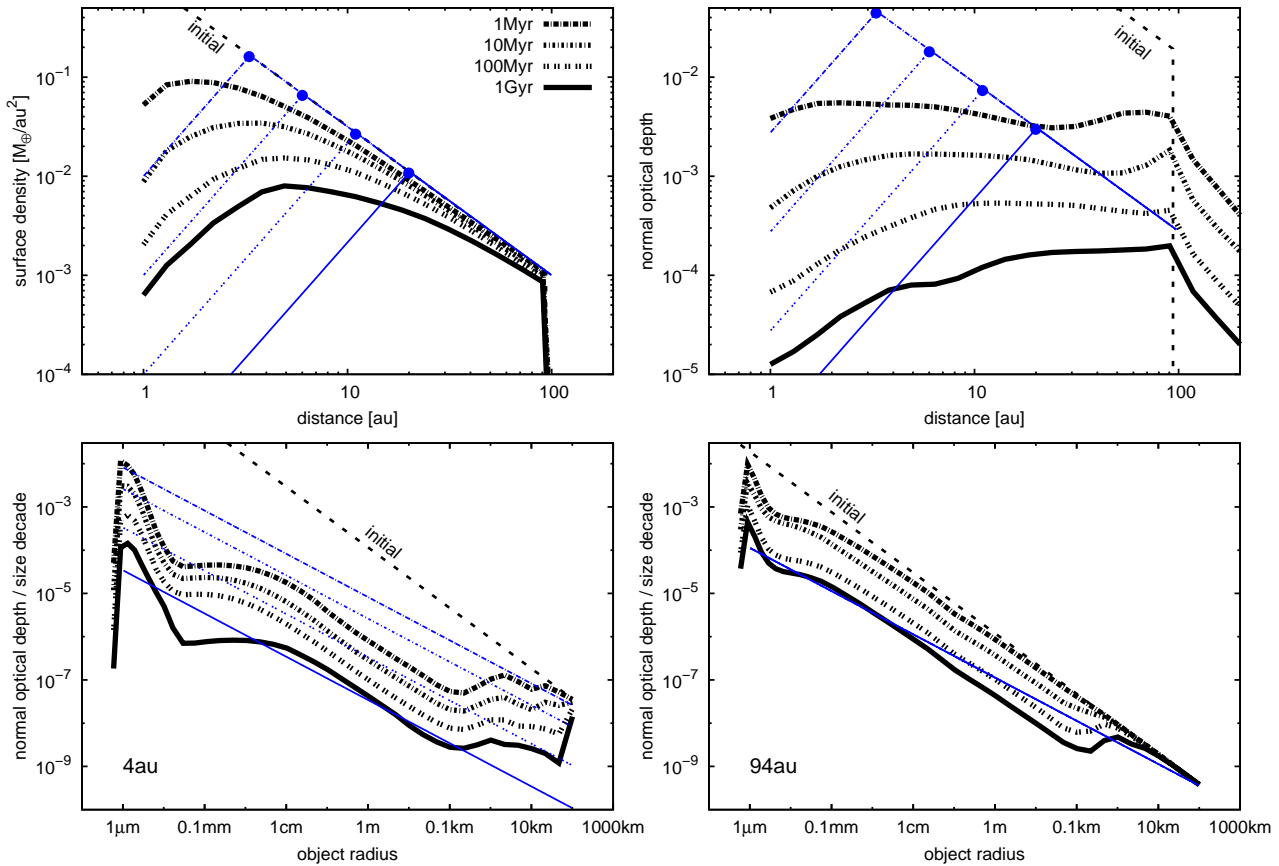} 
\caption{
Radially extended debris disc at different time instants after the 
beginning of the collisional evolution. 
Thick black lines are \texttt{ACE} results,  
thin blue lines represent the model of \citet{Wyatt2007b}. 
\textit{Top:} radial profiles of the surface density \textit{(left)}
and the normal optical depth \textit{(right)}.
The blue dots on the straight lines indicate the distances where the collisional
lifetime of the largest planetesimals equals the current system age.
Beyond these distances the lines turn into the initial profiles. 
\textit{Bottom:} size distributions in terms of optical depths 
at 4\,au \textit{(left)} and 94\,au \textit{(right)}.
At 94\,au, the lines of the \citet{Wyatt2007b} model are the same at  
all time instants, since the collisional depletion of the largest planetesimals 
at that distance has not yet started by 1\,Gyr.
}
\label{fig:ER}
\end{figure*}

\subsection{Extended planetesimal disc}

We first consider the long-term collisional evolution of an initially extended planetesimal disc,
which is a likely remnant of the planetesimal formation process.
In this section, we seek to understand the general features of collisional evolution 
predicted by our simulations.
Therefore, we generated a representative setup by using a set of standard assumptions
that are not yet adjusted to the q$^\mathrm{1}$~Eri system.

We assumed a planetesimal distribution between 1 and 100\,au around a solar-type star, 
having a solid surface density of the standard
Minimum Mass Solar Nebula (MMSN) model: 
\mbox{$\Sigma=1\,\mathrm{M}_\oplus\, \mathrm{au}^{-2} \, (r/\mathrm{au})^{-1.5}$} 
\citep{Weidenschilling1977, Hayashi1981}.
The disc was filled with planetesimals up to 100\,km in radius. 
We set the orbital eccentricities of the planetesimals 
to a mean value of 0.05, in agreement with values expected 
for discs around solar-type stars \citep[e.g.,][]{Pawellek2015}.
Also, smaller bodies were already present within the disc 
at the beginning of the simulation.
The initial size distribution of the bodies followed
$s^{-3.7}$, 
which is the equilibrium distribution of dust
in an infinite collisional cascade
for the material parameters in the strength regime
given in Sect.~\ref{sec:method}  \citep{OBrien2003}.
This particular choice does not really matter, 
since the simulation promptly drives
the system to an equilibrium size distribution that does not depend much on the one 
assumed initially.
This comes with the caveat that the kilometre-sized and larger planetesimals in the outer
disc do not reach a collisional equilibrium by the end of the simulation and
may retain their primordial distribution set by their formation process 
(which is poorly known anyway).
The smallest particles considered in the simulation had radii
of $\approx0.5\,\micron$ that corresponds to the blowout radius, $s_\mathrm{blow}$, 
of silicate grains around a solar-type star.  
Grains smaller than $s_\mathrm{blow}$ were neglected because they are 
pushed on unbound orbits through the direct radiation pressure 
and are expelled from the system on dynamical timescales.

Top panels in Fig.~\ref{fig:ER} depict the evolution of the mass surface density, $\Sigma$, 
and the normal optical depth, $\tau_\perp$, as functions of distance from the star.
While $\Sigma$ reflects the distribution of kilometre-sized, massive bodies, 
$\tau_\perp$ traces the distribution of micron-sized dust grains that dominate the 
cross section.
This difference is best illustrated by considering the region
beyond 100\,au. In this zone, the disc mass surface density is negligible, while the optical 
depth shows a halo that
contains small grains with radii slightly above $s_\mathrm{blow}$.
These grains are in eccentric orbits
with pericentres within the planetesimal zone and apocentres outside.

Bottom panels in Fig.~\ref{fig:ER} depict the size distributions close to the 
inner (4\,au) and outer (94\,au) edge of the planetesimal disc.
The size distributions exhibit a wavy pattern which is more pronounced
towards the dust size range (radii $<$1\,mm), since there are no grains below 
$s_\mathrm{blow}$ that could work as destructive projectiles 
\citep{CampoBagatin1994,Krivov2006,Thebault2007,Wyatt2011}.

Both the radial profiles and the size distributions change with time.
This evolution is described and discussed in detail below, and 
we start with the initial stage. The top right panel of Figure~\ref{fig:ER}
reveals an extremely high initial optical depth ($>10$ at 1\,au),
in the innermost regions of the disc.
It arises because \texttt{ACE} takes the initial slope of the $\Sigma$ profile and
populates the disc initially by small, micron-sized grains in accord with
the $s^{-3.7}$ power law. However, as explained above, that power law tacitly assumes
that the system has an equilibrium distribution of an idealised, infinite collisional
cascade from the very beginning. Obviously, this assumption is not valid for a debris disc
right after its birth. Instead, the initial distribution of dust in a young debris disc
must be the one set by physical processes that operated in the preceding protoplanetary
disc by the time of the gas dispersal.
On any account, \texttt{ACE} gets rid of the ``unphysically'' overabundant
dust particles swiftly and the optical depth drops to the equilibrium values
after a few thousand years.
Already at 1~Myr, the $\tau_\perp$ curve
in Fig.~\ref{fig:ER} is flat, regardless of the profile assumed initially.

The subsequent evolution of the disc depends
on the collisional lifetimes of the disc particles and
can be understood as follows.
Since collisional lifetimes get longer with
increasing particle size and distance from the star,
modifications of the size distributions go from small to large objects
and from inner to outer disc regions.
The small dust that dominates the optical depth depletes and reaches collisional
equilibrium first, larger objects follow.
As a result, at times shorter than the collisional lifetime of the largest objects 
at a given distance, $\tau_\perp$ decreases faster than
$\Sigma$ does.
It is this gradual inside-out, bottom-up transition of the 
size distribution from the initial to the quasi-equilibrium state combined with the initial 
radial distribution that explains
why the radial profiles of $\tau_\perp$ are almost flat, even though the
$\Sigma$ profiles are not.

Once the lifetimes of the biggest
objects are reached at a given distance, the local size distribution keeps its shape,
while $\Sigma$ and $\tau_\perp$ both deplete with their mutual ratio staying constant.
After tens of megayears, material is sufficiently eroded to generate positive slopes
of both $\Sigma$ and $\tau_\perp$ in the inner regions.

After 1\,Gyr, almost all material is reprocessed
and only the largest planetesimals in the outer disc have not yet started
to deplete collisionally and follow their primordial size distribution.
This is seen for objects larger than 10\,km at 94\,au.

\begin{figure*}
\includegraphics[width=0.95\textwidth]{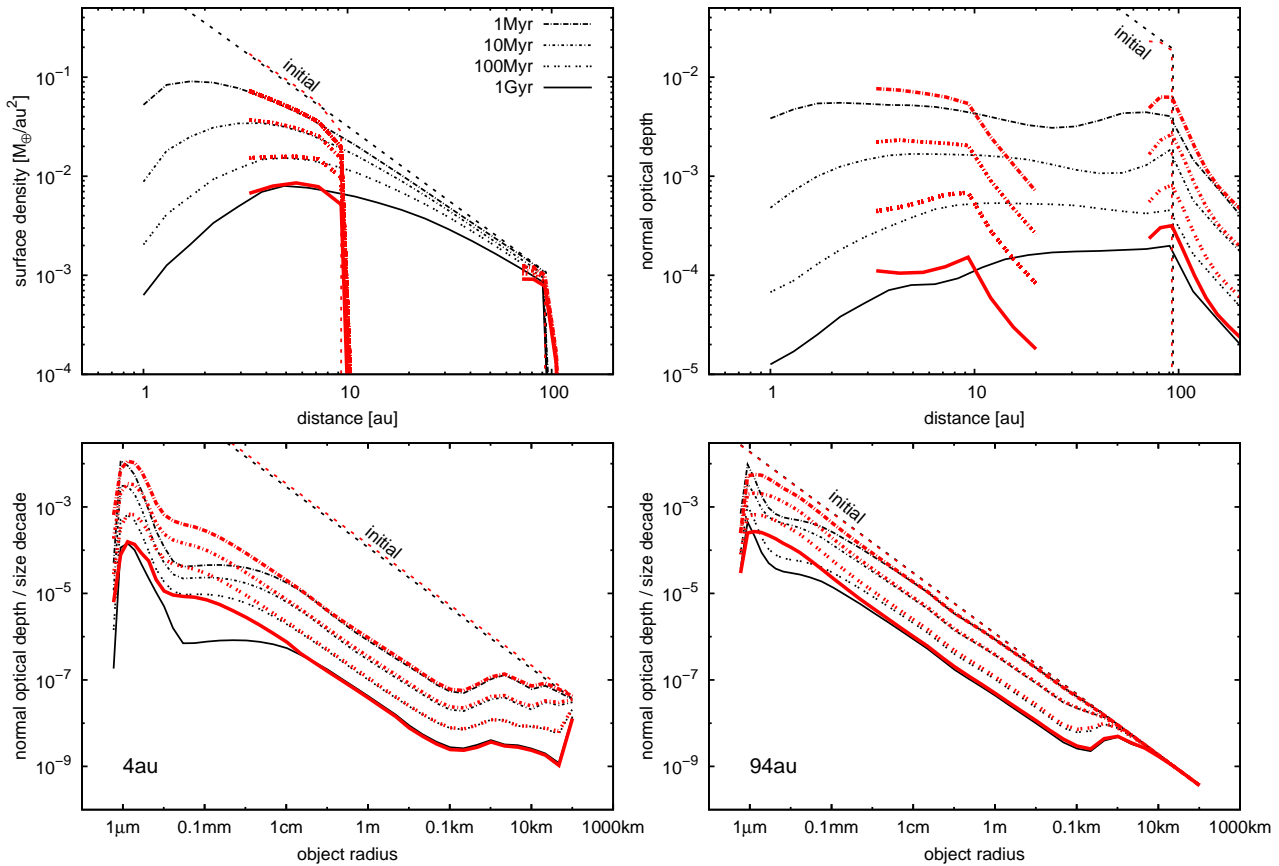} 
\caption{
Evolution of a two-component disc in comparison with that of an extended one.
Thin black lines correspond to the radially
extended planetesimal model presented in Fig.~\ref{fig:ER}. 
Thick red lines show two narrow planetesimal belts, 
between 3--10~au and 70--100~au.
}
\label{fig:ER+2NR}
\end{figure*}

For the sake of comparison,
the analytic steady-state evolution model of \cite{Wyatt2007b} is overplotted in Fig.~\ref{fig:ER}. 
Their model was designed to describe the evolution of a narrow debris ring, but can readily
be applied to an extended debris disc, assuming it to consist of non-interacting concentric annuli
\citep{Kennedy&Wyatt2010}.
To enable direct comparison, we converted the maximum disc mass $M_\mathrm{max}$
and the maximum fractional luminosity of the dust $f_\mathrm{max}$
of \citet{Wyatt2007b} to the surface density and the normal optical depth:
\begin{align}
 \Sigma(r,t) &=  \frac{M_\mathrm{max}}{2\pi\, r\, dr}, \\
 \tau_\perp(r,t) &= 2\frac{r}{dr} f_\mathrm{max}(r,t), 
\end{align}
where $r$ is the radius and $dr$ the width of a planetesimal ring.
We also calculated the evolution of the size distributions 
by means of the equations~(1), (13), and (14) of \cite{Wyatt2007b}.

The parameters of the \citet{Wyatt2007b} model were chosen in such a way that
it matches the \texttt{ACE} run as closely as possible.
Accordingly, all parameters that are shared by the analytic model and the 
\texttt{ACE} simulation were set to the same values.
These include the stellar mass and luminosity,
the initial disc mass,
the eccentricity of the planetesimal orbits,
and the diameter $D_\mathrm{c}$ of the largest planetesimals in the collisional cascade.

There are two important parameters, however, for which the best choice is less obvious.
One is the slope $(2-3q)$ of the size distribution, which is not an input parameter
of the \texttt{ACE} simulation.
We assumed $q=11/6$, or $2-3q = -7/2$.
Another parameter is the specific disruption energy $Q_\mathrm{D}^\ast$, which
must be independent of size in the \citet{Wyatt2007b} model.
We chose $Q_\mathrm{D}^\ast=10,000$\,J/kg, which is 
representative for two different particle populations. 
Indeed, since $Q_\mathrm{D}^\ast$ used in \texttt{ACE} has a V-like shape in a log-log scale 
with a minimum at around 100~m, any constant value above this minimum intersects the 
two branches in the small and large particle size range.
The assumed $10,000$\,J/kg corresponds to the material strength 
of planetesimals with a few tens of kilometres in radius
as well as of millimetre-sized grains.
The latter are important sources for the production of yet smaller dust 
that mainly determines the optical depth. 
Thus, our choice of $Q_\mathrm{D}^\ast$ ensures that 
particles from the lower and upper end of the wide 
size range are taken into account.

As can be seen in Fig.~\ref{fig:ER}, the \texttt{ACE} simulation predicts a significantly
higher mass surface density at later times than the analytic model of \cite{Wyatt2007b}.
This is a consequence of a slower depletion of the planetesimals best visible
in the size distributions at 4\,au.
In contrast to the analytic reference, the \texttt{ACE} simulation involves a 
size-dependent $Q_\mathrm{D}^\ast$. 
The lifetimes of the largest planetesimals are longer, because
their $Q_\mathrm{D}^\ast$'s exceed the value of 10,000 J/kg adopted in the analytic 
model by a factor of several.
Furthermore, the $Q_\mathrm{D}^\ast$'s of the planetesimals
increase with object size as a result of self-gravity.
The latter effect enhances the abundance of larger planetesimals relative
to their disruptive projectiles. 
The slope of the resulting steady-state size distribution of the largest planetesimals
flattens with respect to that in the strength regime or the canonical $s^{-3.5}$ \citep{OBrien2003}.
Thus, at later times, the \texttt{ACE} simulation retains a reservoir of large planetesimals
that is more massive and is able to sustain a given optical depth
or dust mass for a longer period of time than the analytic model.

\subsection{Two narrow planetesimal discs}
\label{sec:2narrow_discs}

We now consider the evolution of a disc whose radial structure 
was additionally sculpted by the formation of planets.
The primary effect of giant planet formation, 
which we only consider here,
is the opening of a planetary gap that splits up the disc.
Since such a gap is already produced before the debris disc phase, 
we can take it into account through 
a suitable choice of the initial density profile.

Accordingly, we cut out
two regions of the extended planetesimal disc presented in the previous section.
The inner disc was assumed to lie between 3 and 10\,au, the outer disc between 70 and 100~au.
Other initial parameters were the same as explained for the extended disc.
In particular, the initial surface density of
both belts followed the MMSN description.
The two belts were then evolved with two independent
\texttt{ACE} runs.

Figure~\ref{fig:ER+2NR} compares the two-component system with the extended disc.
At any given age, the surface densities in the two 
models are very close (Fig.~\ref{fig:ER+2NR}, top left).
There are, however, some deviations between 
the shape of the size distributions in the two models, especially
in the outer zone (Fig.~\ref{fig:ER+2NR}, bottom right).
These deviations are related to the collisional interaction of different disc regions.
In the extended disc model, the dust grains with sizes just above the blowout
size, produced at a certain distance,  
are sent by radiation pressure to eccentric orbits, and therefore,
enter the outer disc zones where they collide with the local material.
This causes an overabundance of grains just above the blowout limit, which
enhances the depletion of larger grains.
This effect favours the formation of distinct dips in the size distributions, as  
seen 
around 5\,\micron{} at 94\,au.
Since the two-component model simulates the two belts independently,
such a collisional interaction is neglected, and so,
the dips are absent.
As a consequence, the planetesimal rings of the two-component model 
tend to have somewhat higher optical depths than in the extended disc simulation
(Fig.~\ref{fig:ER+2NR}, top right).

We emphasize that our model treats the evolution of the inner
and outer planetesimal belts self-consistently since both
originate from a common initial surface density profile. 
Possible interactions with one or more planets within the gap that 
separates the two components are not directly included.
However, such interactions can be taken into account through the 
choice of initial disc parameters.
For instance, the gap width reflects the number and mass of planets.

\section{Application to \lowercase{q}$^1$~Eri}
\label{sec:application}

\subsection{Star}
\label{sec:stellar_properties}

We took stellar parameters   
from the \textit{Herschel}/DUNES final archive \citep{Eiroa2013}. 
The star has the spectral type F8\,V, 
a luminosity of 1.5\,L$_\odot$, 
an effective temperature of $6155\,$K, a mass of 1.12\,M$_\odot$, 
a metallicity of $[\mathrm{Fe/H}]=-0.04$, and a 
surface gravity of $\log(g)=4.48$ (CGS).
The star is at a distance of 17.4~pc \citep{vanLeeuwen2007}. 

We adopted the stellar photosphere model from \cite{Eiroa2013}, 
which has been obtained by an interpolation 
in the PHOENIX/GAIA model grid \citep{Brott&Hauschildt2005} with a
normalisation to optical and \textit{WISE} fluxes.

Age estimations of q$^\mathrm{1}$~Eri span a wide range, from 0.3 to 4.8~Gyr 
\citep[][and references therein]{Liseau2008}.
However, the strength of chromospheric activity and the luminosity in X-rays 
suggest that the star is $\gtrsim1\,$Gyr old.
In this paper, we assume an age of around 1~Gyr.

\subsection{Observational data}

Photometric data have been collected from the 
\textit{Herschel}/DUNES archive\footnote{http://sdc.cab.inta-csic.es/dunes/jsp/homepage.jsp} 
and the literature.
From the set of available data, we use the measurements 
that are at least $1\sigma$ above 
the stellar photosphere model, thus
indicating the infrared excess (Table~\ref{tab:fluxes}).
A \textit{Spitzer}/IRS spectrum, reduced by the c2d Legacy Team Pipeline,
was also taken from the DUNES archive.

In addition, we extracted radial surface brightness profiles from the
\textit{Herschel}/PACS images provided in the DUNES archive. At 70, 100, and
160~\micron, cuts were performed along the minor and major axes of the observed
disc, with position angles of 144\degr{} and 54\degr{}
east of north, respectively. 
The stellar position was assumed to coincide with the
intensity-weighted centroid of the disc. 
Profiles of the minor and major axes were obtained by averaging the two sides of the disc.
Uncertainties in intensity are
dominated by differences between opposing branches of both axes, added in
quadrature to a minor contribution from background noise that was estimated from
an annular aperture. The resulting profiles and error bars are depicted in
Fig.~\ref{fig:SED+profiles}. Horizontal error bars reflect the bin widths in the
underlying images: 1\arcsec{} at 70 and 100~\micron, 2\arcsec{} at 160~\micron.

\begin{table}
\caption{Photometry of the q$^\mathrm{1}$~Eri disc.}
\label{tab:fluxes}
\begin{tabular}{cccc}
\hline\\[-2ex]
Wavelength  & Flux density   &Instrument & Ref.\\
 $[\micron]$ & [mJy]  &  &             \\
\hline\\[-2ex]
18      & $315.3\pm39.7$    & \textit{AKARI}          & (1) \\
22      & $218.1\pm4.0$     & \textit{WISE}           & (1) \\
24      & $184.8\pm3.8$     & \textit{Spitzer}/MIPS   & (1) \\
25      & $201.1\pm26.1$    & \textit{IRAS}           & (1) \\ 
60      & $617.4\pm80.3$    & \textit{IRAS}           & (1) \\
70      & $896.2\pm26.9$    & \textit{Herschel}/PACS  & (1) \\
90      & $904\pm60$        & \textit{AKARI}          & (4) \\
100     & $897.1\pm26.9$    & \textit{Herschel}/PACS  & (1) \\
160     & $635.9\pm31.8$    & \textit{Herschel}/PACS  & (1) \\
250     & $312.3\pm25.6$    & \textit{Herschel}/SPIRE & (1) \\ 
350     & $179.9\pm14.6$    & \textit{Herschel}/SPIRE & (1) \\
500     & $78.4\pm9.8$      & \textit{Herschel}/SPIRE & (1) \\
850     & $30.3\pm4.1$      & JCMT/SCUBA-2            & (6) \\
870     & $39.4\pm4.1$      & APEX/LABOCA             & (3) \\
1200    & $<17$             & SIMBA/SEST              & (2) \\
6800    & $0.093\pm0.017$   & ATCA                    & (5) \\
\hline
\end{tabular}\\
\textbf{References.}
(1) Based on data from the DUNES Archive at Centro de Astrobiolog\'ia \citep{Eiroa2013}. 
(2) \citet{Schuetz2005}.
(3) \citet{Liseau2008}. 
(4) \citet{Yamamura2010}. 
(5) \citet{Ricci2015}. 
(6) Holland et al., in preparation.
\end{table}

\subsection{Two-component model}

As explained in Sect.~\ref{sec:intro}, the q$^1$~Eri system shows strong indications for 
harbouring a two-component debris disc. 
We performed additional tests to confirm that the presence of a warm component is certain. 
One test was to vary the effective temperature of the star by $\pm100$\,K, which is the scatter
in the literature values \citep{Nordstroem2004, Chen2006}.
For each stellar temperature, we generated a photospheric model and then fitted
two-component disc models to the SED, 
assuming the dust distributions to follow power laws.
The warm components we inferred were moderately different.
However, in all the cases a warm component was clearly needed to explain the data.
Another test was to overplot the q$^\mathrm{1}$~Eri system in 
figure~5 of \cite{Kennedy&Wyatt2014} that presents a $\chi^2$ 
map calculated from fitting a single-temperature blackbody (i.e.,
a one-component disc) to fiducial two-temperature (i.e., two-component) discs.
We find q$^\mathrm{1}$~Eri to lie close to the contour line  
$\chi^2_\text{red} \approx 20$, meaning that this system is highly inconsistent with 
a one-component structure.
Finally, we estimated an uncertainty in the normalisation of the photospheric 
model to the optical and \textit{WISE} fluxes to be about 2\%. A two-component fitting
similar to that described above, but with the photospheric model made by 2\% brighter, 
also showed that the warm component was needed to reproduce the SED.

Given these results, we applied the \texttt{ACE}-based two-component 
model presented in Sect.~\ref{sec:2narrow_discs} 
to explain the SED and the \textit{Herschel} 
surface brightness profiles of the q$^\mathrm{1}$~Eri system.
We assumed the stellar properties as described in Sect.~\ref{sec:stellar_properties} and 
simulated a set of several two-component models 
where we varied the following parameters:
the location, width, and initial mass (in terms of the MMSN multiples) of the inner and outer components, 
the eccentricities of the planetesimal orbits, the size of the largest planetesimals, and the
chemical composition of the disc particles.
For the purpose of surface profile modelling, we generated 
synthetic images from the \texttt{ACE} models 
at 70, 100, and 160\,\micron{} with a position 
angle of 54.2\degr{} (east of north), 
which is the mean value seen in the three \textit{Herschel}/PACS images.
The synthetic images were then convolved with $\alpha$~Boo PSFs. 
The PSFs were rotated by 83\degr{} counter clockwise to align 
the maps with the telescope pupil orientation on the sky 
during the q$^\mathrm{1}$~Eri observations.

Fitting the model to the observational data is
challenging, because we can only vary the parameters of the parent planetesimals 
and not the dust distributions. 
The impact of  parameter variations on the resulting dust distributions, and thus, 
on the thermal emission properties is difficult to predict quantitatively.
Several time-consuming trial-and-error \texttt{ACE} simulations are necessary 
to change the dust distribution in a desired way.
Furthermore, different combinations of disc parameters 
can produce similar SEDs and brightness profiles, 
and therefore, a unique solution does not exist.
We aimed at finding one plausible model that agrees with the observations reasonably well.
We probed various maximum orbital eccentricities of planetesimals (0.03, 0.05, and 0.1)
and their largest radii (50 and 100\,km).
Inner and outer edges of the inner planetesimal belt
were varied between 3 and 20\,au, and those of the outer belt between 65 and 135\,au.
We assumed dust particles to be a mixture of astrosilicate \citep{Draine2003b}
and water ice \citep{Li1998}
and chose the volume fraction of ice to be a free parameter.
Ice fractions of 0, 50, 70, and 90 per cent were explored.

\begin{figure}
\includegraphics[width=\columnwidth]{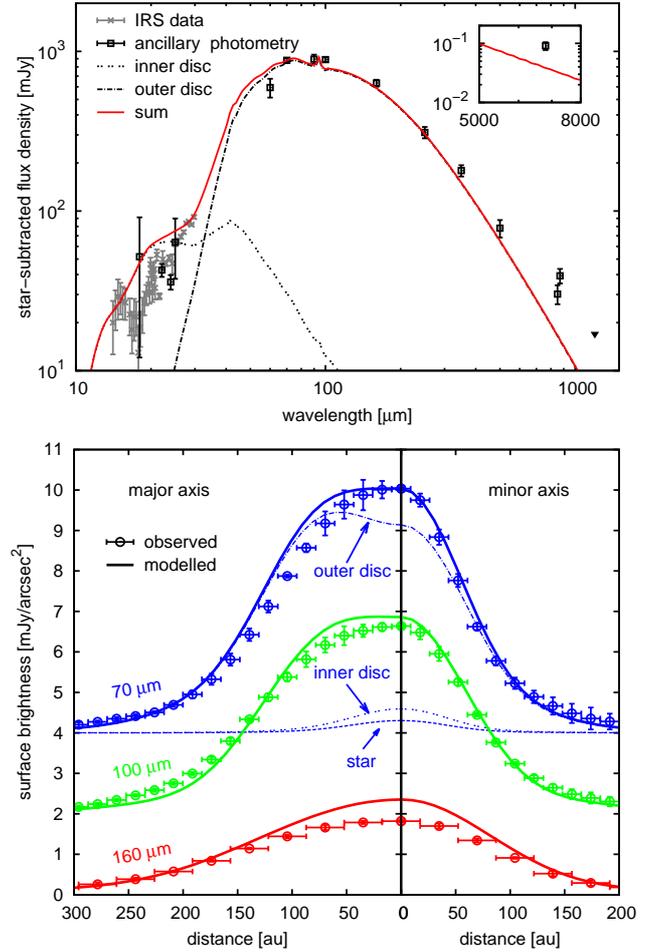}
\caption{Thermal emission of a two-component model
that we found in satisfying agreement with the observational data (dots).
\textit{Top:} the full SED (solid) along with contributions 
from the inner (dotted) and outer (dashed-dotted) component. 
The inset in the upper right corner shows how the model matches the ATCA measurement at 6.8\,mm.
\textit{Bottom:} surface brightness profiles across the minor and 
major axes for 70 (blue), 100 (green), and 160\,\micron{} (red).
For 70\,\micron{}, contributions of the star (dashed line), inner component (dotted),
outer component (dashed-dotted), and the sum of all three (solid) are shown.
At 100 and 160\,\micron{}, only the total profiles
are depicted since they come almost entirely from the outer component.
For better visibility, the 70\,\micron{} and 100\,\micron{} profiles were shifted vertically
by constant values of +4 and +2\,mJy/arcsec$^2$, respectively.
}
\label{fig:SED+profiles}
\end{figure}

\begin{table*}
\caption{Parameters of the two-component model shown in Fig.~\ref{fig:SED+profiles}.}
\label{tab:parameters}
 \begin{tabular}{lccccccccc}
\hline\\[-2ex]
Region      & Range$^\mathrm{a}$   & $s_\mathrm{max}^{\phantom{\mathrm{max}}\mathrm{b}}$ & $e_\mathrm{max}^{\phantom{\mathrm{max}}\mathrm{c}}$   
& Material$^\mathrm{d}$ & Incl.$^\mathrm{e}$ & Optical depth$^\mathrm{f}$ & $M_\mathrm{dust}^{\phantom{\mathrm{dust}}\mathrm{g}}$  & $M_\mathrm{disc}^{\phantom{\mathrm{disc}}\mathrm{h}}$     & Age$^\mathrm{i}$     \\
            & [au]    & [km]             &                    &                              & [\degr] &                      & [M$_\oplus$]       & [M$_\oplus$] & [Gyr]   \\
\hline\\[-2ex]
 Inner       & 3--10   & \multirow{2}{*}{50}   & \multirow{2}{*}{0.05}  & \multirow{2}{*}{30\%\,silicate + 70\%\,ice}   & \multirow{2}{*}{70} & $1\times10^{-4}$     & $1.5\times10^{-5}$ & 3.4  (95)    & \multirow{2}{*}{1.3}  \\
 Outer       & 75--125 &                       &                        &                                               &                     & $1\times10^{-3}$     & $1.7\times10^{-2}$ & 81.7 (167) &                       \\ 
   \hline
\end{tabular}\\
\raggedright
\textbf{Notes.}
$^{\mathrm{(a)}}$~Radial extent of the planetesimal belts; 
$^{\mathrm{(b)}}$~radius of the largest planetesimals; 
$^{\mathrm{(c)}}$~maximum eccentricity of the planetesimal orbits; 
$^{\mathrm{(d)}}$~dust material; 
$^{\mathrm{(e)}}$~disc inclination from face on;
$^{\mathrm{(f)}}$~optical depth at the centre of the planetesimal belts (at the end of the simulation);
$^{\mathrm{(g)}}$~dust mass including grains up to 1\,mm in radius (at the end of the simulation); 
$^{\mathrm{(h)}}$~initial mass in parentheses and end mass of the whole planetesimal disc 
(bodies up to $s_\mathrm{max}$ in radius);
$^{\mathrm{(i)}}$~simulation time.
\end{table*}

We finally achieved the model presented in Fig.~\ref{fig:SED+profiles}.
The corresponding parameters are summarised in Table~\ref{tab:parameters}.
The model profiles overpredict the 160\,\micron{} emission in the central region
by $\approx20$~per cent. This can reflect some inaccuracy of the collisional model. 
Also, the modelled SED predicts submm and mm flux densities
to be by a factor of 2--3 lower than those measured by LABOCA, SCUBA-2, and ATCA. 
Since the images from SCUBA-2 (Holland et al, in prep.) and LABOCA \citep{Liseau2008} 
are extended eastwards, 
the most likely reason for the discrepancy is a background galaxy. 
The flux densities plotted are aperture fluxes.
The peak fluxes~-- for instance, $20.1 \pm 2.7$\,mJy/beam of the SCUBA-2 measurement~-- is 
more in line with the model prediction.
Overall, we deem the resulting model satisfactory for the purposes of this paper.

The initial surface density of both components corresponds to 5.3 times the MMSN model.
The position of the outer component is consistent with the ring-like disc found through
a deconvolution of the \textit{Herschel} 100\,\micron{} image by \cite{Liseau2010}.
They quote a surface brightness maximum at 85\,au and a belt width of about 35 to 45\,au. 
The location of the inner disc is beyond the orbit of the reported RV planet HD~10647b. 
Although there is a large gap between both components, the overall synthetic 
surface brightness profiles show no signs of a dip and are close to the observed ones.
The best disc inclination is 70\degr{} (from face on), which is 
close to 76\degr, 
the value inferred from the HST scattered light image (K. Stapelfeldt, priv. comm.).

The planetesimal orbits have eccentricities \mbox{$\leq0.05$}, which implies relatively
low impact velocities in the disc and a rather depressed production of small dust.
However, planets that reside within the disc can push 
planetesimals to more eccentric, and also inclined, orbits 
through secular perturbations \citep{Mustill2009}.
In particular, this planetary stirring is likely to operate in the inner component, 
which is close to HD~10647b.
There is also the possibility that further, yet undiscovered, 
planets in the gap between both planetesimal belts enhance  
the eccentricities in the outer component. 
Higher values of the eccentricity are not excluded by our modelling 
since there are degeneracies between the eccentricity 
and other model parameters (see Sect.~\ref{sec:discussion}).

The dust composition determines the grain temperatures and emission properties.
During the modelling process, we noticed that the rise of the SED in the mid infrared 
shows a strong dependence on the ice fraction.
This allows us to improve the agreement with the characteristic bump (``shoulder'') that is
exhibited by the IRS spectrum in the 10 to 30\,\micron{} range.
Larger fractions of ice reduce the emission of the inner component, 
whereas the one of the outer component remains nearly unchanged.
This is due to the fact that the mid-infrared emission mostly comes from small grains (1-10\,\micron{} radius),
while the far-infrared emission is produced by larger grains (\mbox{$>10$\,\micron}) 
with an emissivity that is not strongly dependent on the chemical composition.
The higher the ice fraction, the steeper the rise of the total SED.
However, the distinct silicate feature around 18\,\micron{} is still well pronounced
and is able to reproduce the bump seen in the IRS spectrum. 
A mixture of 30~per cent silicate
and 70~per cent water ice for inner and outer component provides a good overall 
match to the SED and the surface brightness profiles.

\section{Discussion}
\label{sec:discussion}

To verify the scenario considered here, it will be crucial to resolve 
the proposed inner belt. 
At 0.17\arcsec-0.57\arcsec, the inner belt would be difficult to see with \textit{Hubble} due to the 
large inner working angle. 
The latest extreme adaptive optics instruments allow us to image much closer to the star. 
For instance, SPHERE/IRDIS on the VLT has an inner working angle of 0.15\arcsec{} in its standard mode, 
opening up these regions to scattered light observations for the first time.
We performed a test by generating scattered light radial profiles of the inner belt at 1.6\,\micron.
We compared the profiles with the IRDIS detection limit 
provided by the SPHERE Exposure Time Calculator\footnote{https://www.eso.org/observing/etc/}.
The surface brightness along the major axis, \mbox{$\sim0.2$\,mJy/arcsec$^2$}, 
is about three orders of magnitude below the detection limit.
For SPHERE/ZIMPOL with a smaller working angle at 0.55\,\micron, 
the inner belt could be traced along the minor
axis where it appears brighter in polarized light, \mbox{$\sim4$\,mJy/arcsec$^2$}. 
However, even in that case we are still one order of magnitude below the detection limit.
This indicates a big challenge in resolving the inner belt with present-day observational facilities.
However, the coming \textit{JWST} might offer better chances for such a detection.

The proposed two-component scenario also implies a set of alleged planets between the inner and
outer belts. The number of planets is unknown, as are their orbits and masses.
Nevertheless, rough estimates can be made by assuming that the presumed planets are
in nearly-circular, nearly-coplanar orbits with a uniform logarithmic spacing of
semi-major axes \citep{Faber&Quillen2007}. The set of planets should be tightly
packed dynamically to ensure that no dust-producing planetesimal belts have survived
between the orbits, while the planets themselves should be at least marginally stable
against mutual perturbations over the age of the system of $\sim1$\,Gyr. Using equations~(3)
and (6) of \citet{Faber&Quillen2007}, we estimate that the cavity between $\sim10$\,au and
$\sim80$\,au could be populated by 4--5 planets of at least Neptune mass, with
a semi-major axis ratio between neighbouring planets of $\sim1.6$ (i.e., similar to
the orbital radius ratio of Earth and Mars in the Solar System). The minimum total
mass of these planets would amount to $\sim0.2$ Jupiter masses.

There are currently two planetary systems known to have an architecture similar to the one proposed for
q$^1$~Eri. One is our own Solar System. Another is a prominent system around HR~8799, with its
four directly imaged massive planets \citep{Marois2008,Marois2010}, an outer
massive ``Kuiper belt'' exterior to and an inner debris belt interior to the planetary orbits
\citep[see, e.g.,][]{Reidemeister2009,Su2009}.
However, there are also fine differences between these cases. The q$^1$~Eri system is known to contain a 
close-in Jupiter-mass radial velocity planet whereas our Solar System harbours four terrestrial planets.
Around HR~8799, no planets have been discovered  so far inside the inner dusty belt.
It is interesting to note that the known planet of q$^1$~Eri system is orbiting at $\sim2$~au, whereas
the proposed inner belt must be located farther out, at 3 to 10~au. Thus, the
belt was likely located outside the snow line when it formed which, 
assuming an ice sublimation temperature at the protoplanetary phase of 170~K \citep{Lecar2006},
must have been at $\sim3.3$~au from the star.
This means that the planet could have comfortably formed in the zone interior to the ``asteroid belt''.
No migration through the belt is required here~-- unlike in the systems such as 
HD~69830 where the planets and the belt are all located inside the snow line
\citep{Payne2009}.

Due to numerous degeneracies, the model found in our study is by far not unique.
For instance, we cannot constrain tightly a set of gas or icy giants that separates the 
two components, and consequently, it is unknown to what degree these would stir the components.
However, we expect that at least the inner component is rather strongly affected by the 
known nearby giant planet HD~10647b which is on a close-in ($a_\mathrm{pl}\approx2$\,au), 
eccentric orbit ($e_\mathrm{pl}\approx0.2$).
Equation~(6) of \cite{Mustill2009} yields a forced eccentricity of $\sim0.1$ that could be imposed
by HD~10647b on planetesimals at 5\,au. 
Note that this is merely a rough estimate since only one planetary perturber is assumed, but
it shows that the inner planetesimal disc may have a distinctly higher level of dynamical excitation
than assumed in our model.
If so, there would be a faster collisional depletion, resulting in a lower 
amount of material in the inner disc.
However, the size of the largest planetesimals and their critical fragmentation energy
are not well constrained, too.
If there are larger and harder planetesimals within the disc, the collisional depletion would be slower, 
which might ensure that the inner system contains sufficient dust after $\approx1$\,Gyr to explain
the observations.  
The collisional timescales can also be prolonged if the inner disc were located farther out and/or had a 
larger width.

Furthermore, the slope of the assumed initial density profile
of a protoplanetary disc is completely uncertain \citep[e.g.,][]{Kuchner2004,Raymond2005,Chiang&Laughlin2013}.
Steeper profiles are preferred for in-situ formation scenarios of close-in giant planets, 
where the disc mass is more centrally concentrated.

The chemical composition of the dust is another crucial point.
In our model, the dust contains 30~per cent silicates and 70~per cent water ice 
in inner and outer component.
However, the inner component lies between 3 and 10~au.
Since, in contrast to protoplanetary discs,
the gas pressure in debris discs is negligible, ice sublimates at lower temperatures of
$\sim100$\,K \citep{Kobayashi2008}, suggesting the ice line for blackbody grains to lie
at $\sim10$~au.
Thus the presence of icy grains in the inner component is questionable.
Other materials, e.g. porous silicate, for the inner component may be a better choice. 
Such a scenario agrees, at least qualitatively, with the compositional model of \cite{Tanner2009}.

\section{Conclusions}
\label{sec:conclusions}

We consider a scenario that might naturally explain the two-component structure
observed in many debris discs. 
This includes the formation of giant planets in an initially extended planetesimal disc
that is still immersed in gas (protoplanetary phase).
The nascent planets forming within the disc open the gap around their orbits 
by scattering the planetesimals away.
Several planets together can generate a single broad gap and, as a consequence, 
the initially extended disc splits up into an inner and an outer region. 
After the gas dispersal, both planetesimal discs start to evolve collisionally.
Even though the dust produced in collisions between planetesimals is spread over large distances
by radiative and corpuscular forces,
the structure of the underlying planetesimal distribution is that of a two-component disc.
In this paper, a detailed collisional model involving realistic physics
is used to simulate the long-term collisional evolution of a two-component disc 
around a solar-type star.
Our conclusions are as follows:

\begin{enumerate}
\item[1.] 
From a comparison of the collisional simulations with predictions 
of a simplified analytic model of \cite{Wyatt2007b}, 
we show that the results are quite different. 
The surface density provided by the collisional simulations reveals much more material 
close to the star after an evolution period of 1\,Gyr. 
The optical depth does not rise as steeply with distance from the star as predicted by the 
analytic model.
This means that the discs evolving in a steady-state regime may be 
able to retain larger amounts of material in the inner region at older ages than thought before.

\item[2.] That scenario implies the presence of as yet undiscovered planets. 
Since it is the only scenario proposed so far for systems without significant transport 
(which is the case for stars of earlier spectral types), 
and since there are many two-component discs around such stars, 
this suggests the Solar System-like architecture with sets of planets 
between the Kuiper belts and asteroid belts may be quite common.
If no planets populate the gap between both components, the only remaining explanation 
for the formation of a two-component disc will be 
that planetesimals, for whatever reason, did not form in a certain region of the disc.

\item[3.] 
The two-component model is able to reproduce the thermal 
emission of the q$^1$~Eri disc over a wide wavelength range. 
This is achieved
with an inner planetesimal belt
between 3 to 10\,au and an outer one between 75 to 125\,au.
The initial surface density of both components is approximately 
five times the Minimum Mass Solar Nebula model.
The belts contain objects up to 50\,km in radius
and the planetesimal orbits have a maximum eccentricity of 0.05.
Silicate grains with 70~per cent ice content are assumed to reproduce the thermal emission properties.
Due to a number of degeneracies, this is not a unique solution. 
To check the scenario itself and break some of the degeneracies, 
it would be vital to resolve the inner belt. 
There must also be a chance in the future to detect the expected planets.
\end{enumerate}

\section*{Acknowledgements}
We thank Karl Stapelfeldt for sharing with us the radial brightness profile derived from the
scattered light observations of q$^1$~Eri,
Nicole Pawellek for helpful discussions on SED fitting, and Mark Wyatt for commenting on the 
manuscript draft.
We also thank the referee for insightful comments that helped to improve the paper 
significantly.
A.V.K., T.L., F.K., and S.W. acknowledge the support by the 
\textit{Deut\-sche For\-schungs\-ge\-mein\-schaft} (DFG) through projects 
\mbox{Kr~2164/13-1},
\mbox{Kr~2164/15-1},
\mbox{Lo~1715/2-1}, and
\mbox{Wo~857/15-1}.
M.B. acknowledges support from a FONDECYT Postdoctral Fellowship, project no.~3140479.




\bibliographystyle{mnras}

\bsp	
\label{lastpage}
\end{document}